\documentclass[journal]{IEEEtran}
\usepackage{amsmath,amsfonts}
\usepackage{algorithmic}
\usepackage{array}
\usepackage[caption=false,font=normalsize,labelfont=sf,textfont=sf]{subfig}
\usepackage{stfloats}
\usepackage{verbatim}
\usepackage{graphicx}
\usepackage{epsfig}
\usepackage{tikz}
\newcommand*\numrounded[1]{\tikz[baseline=(char.base)]{
            \node[shape=circle,draw,inner sep=0.7pt] (char) {#1};}}

\usepackage{booktabs}
\usepackage{color}
\usepackage{tabularx}
\usepackage[style=ieee,backend=biber]{biblatex} % load biblatex instead of cite
\usepackage{hyperref} % should go after biblatex
\hypersetup{
    colorlinks=true,
    linkcolor=blue,
    filecolor=magenta,      
    urlcolor=cyan,
}
\hypersetup{linkcolor=black}
\hypersetup{citecolor=black}
\usepackage{multirow}

\usepackage{amssymb} % for checkmark symbols
\usepackage{pifont}  % for xmark if needed

\usepackage[table]{xcolor}

\newcommand{\cmark}{\checkmark} % checkmark
\begin{document}

\title{Network Self-Configuration based on Fine-Tuned Small Language Models}

\author{\IEEEauthorblockN{Oscar G. Lira\IEEEauthorrefmark{1}, Oscar M. Caicedo\IEEEauthorrefmark{2}, Nelson L. S. Da Fonseca\IEEEauthorrefmark{1}}
\thanks{O. Lira and N.L.S. Da Fonseca are with the Institute of Computing, University of Campinas, Campinas 13083-852, Brazil. E-mail: o224415@dac.unicamp.br, nfonseca@unicamp.br.}
\thanks{O. M. Caicedo is with the Departmento de Telematica, Universidad del Cauca, Popay\'an, 19001, Colombia. E-mail: omcaicedo@unicauca.edu.co.}
}

% The paper headers
\markboth{IEEE XXXX, ~Vol.~XX, No.~XX, XXXX~2025}%
{Lira and da Fonseca: A Network Configuration Approach based on Fine-Tuned Small Language Models}

\maketitle

\begin{abstract}
As modern networks continue to expand in scale and complexity, manual configuration processes become increasingly inefficient and susceptible to human error. Although intent-driven self-configuration using large language models has demonstrated considerable potential, these models remain computationally intensive and resource-hungry, and they often raise privacy concerns due to their reliance on external cloud infrastructure. This work presents SLM\_netconfig, a fine-tuned, small-language-model–centered framework that adopts an agent-based architecture and employs parameter-efficient adaptation techniques to translate configuration intents—expressed as natural-language requirements or questions—into syntactically and semantically correct network configurations. The system is trained using a domain-specific dataset generated through a dedicated pipeline grounded in vendor documentation, ensuring strong alignment with real-world configuration practices. Extensive evaluation demonstrates that SLM\_netconfig, when operating with its question–configuration model, achieves higher syntactic accuracy and goal accuracy than LLM-NetCFG, while significantly reducing translation latency and producing concise, easily interpretable configurations. These results indicate that fine-tuned small language models, as embodied in SLM\_netconfig, can provide efficient, accurate, and privacy-preserving automated configuration generation entirely on-premise, establishing them as a practical and scalable solution for modern autonomous network configuration.
\end{abstract}


\section{Introduction}
The increasing complexity of network configurations and the growing demand for automated network management have created an urgent need for efficient, accurate methods to generate autonomous device configurations. In this context, the Zero-touch Network and Service Management (ZSM) framework, introduced by the European Telecommunications Standards Institute (ETSI), aims to achieve self-operating, self-maintaining, and self-optimizing networks with minimal human intervention, thereby transforming the landscape of network automation \cite{ETSI-ZSM}. A fundamental component of ZSM is self-configuration, which enables networks to set and adjust operational parameters autonomously. However, within the ZSM paradigm, the autonomous generation of syntactically valid and semantically accurate configurations that align with network intents remains an open challenge.

The use of Large Language Models (LLMs) represents a promising approach for automating configuration generation within Intent-Driven Networking (IDN) frameworks \cite{Yan@Sma, Ang@Lln}. Leveraging their ability to process and generate natural language, LLMs can translate high-level business or technical intents into executable configuration commands. ETSI’s intent-based management framework formalizes this concept by defining a transformation process from high-level intents to enforceable policies, thereby enabling networks to adapt dynamically to operational requirements. Nevertheless, current LLM-based systems \cite{NETBUDDY,GeNet,Wha@Mon,Jeo@Swi,don2024llm,LLM@Mek} that convert natural language intents into network commands face limitations in terms of high computational costs, restricted data privacy, and insufficient accuracy. These limitations stem primarily from reliance on external computing facilities and LLMs, the size of LLMs, which often comprises billions and even trillions of parameters, and the lack of network-specific fine-tuning. As a result, they are less effective in generating configurations that accurately reflect network goals. In \cite{Osc@Lar}, we introduced LLM-NetCFG, a locally deployed LLM designed to mitigate these challenges. However, LLM-NetCFG still exhibits hallucination effects, which negatively impact both accuracy and translation latency.

A critical issue in LLM-based network configuration generation is hallucination, in which a model produces syntactically correct but semantically invalid or irrelevant outputs. In autonomous configuration, hallucination manifests as commands that diverge from the intended goal, contain unsupported syntax, or reference non-existent network entities. Such inaccuracies undermine the reliability and safety of intent-driven automation, as even minor configuration errors may cause service interruptions or security breaches. The root causes of hallucination lie in insufficient domain-specific grounding during model training and the scarcity of structured, high-quality datasets that tightly couple natural language intents with verified configurations. Consequently, mitigating hallucination requires not only architectural and fine-tuning improvements but also robust dataset engineering practices that ensure semantic consistency between intents and configuration outputs.

Prompt engineering and fine-tuning are complementary strategies for adapting language models to domain-specific tasks and reducing hallucinations. Prompt engineering focuses on crafting structured instructions or templates that condition a model’s reasoning and outputs without modifying its parameters. This approach enables rapid, resource-efficient specialization; however, its performance is highly dependent on prompt quality, and it provides limited guarantees of determinism or domain-grounded responses. Fine-tuning, in contrast, adjusts a model’s internal parameters using curated, domain-specific datasets, enabling the model to internalize specialized knowledge and reduce hallucinations. While this strategy offers improved accuracy and stronger contextual alignment, it requires substantial computational resources and high-quality annotated data.

%\omc{Prompt engineering and fine-tuning, and retrieval-augmented generation (RAG) constitute three complementary strategies for adapting language models to domain-specific tasks and reducing hallucinations. Prompt engineering focuses on crafting structured instructions or templates that condition a model’s reasoning and outputs without modifying its parameters. This approach enables rapid, resource-efficient specialization; however, its performance is highly dependent on prompt quality, and it provides limited guarantees of determinism or domain-grounded responses. Fine-tuning, in contrast, adjusts a model’s internal parameters using curated domain-specific datasets, enabling the model to internalize specialized knowledge and reduce hallucinations. While this strategy offers improved accuracy and stronger contextual alignment, it requires substantial computational resources and high-quality annotated data. RAG enhances model performance by dynamically retrieving relevant external information from a knowledge base during inference, enabling responses grounded in factual data without requiring parameter modification; however, its reliability depends on retrieval quality and database maintenance.}

This paper introduces SLM\_netconfig, a hybrid approach combining Fine-Tuned Language Models (FLMs) and Small Language Models (SLMs) to enhance the efficiency and accuracy of network self-configuration. By adapting pretrained LLMs to domain-specific tasks without re-training all parameters, the proposed approach significantly reduces computational overhead, memory footprint, and storage requirements while maintaining or improving accuracy and data privacy. SLM\_netconfig comprises a modular architecture that includes (i) a fine-tuning process for small configuration models, (ii) a dataset generation pipeline for extracting configurations from technical documentation, and (iii) two fine-tuned small models, requirement-to-configuration and question-to-configuration, that autonomously perform specialized configuration tasks.

SLM\_netconfig combines prompt engineering to structure the model’s interactions and guide the reasoning process for requirement interpretation and configuration synthesis. Fine-tuning aligns the model’s internal representations with the syntactic, semantic, and intent-configuration dependencies characteristic of network management. This combination enables the generation of syntactically valid and semantically coherent configurations while mitigating hallucinations, addressing the limitations of external LLMs, and achieving an effective balance between efficiency, accuracy, and privacy in autonomous network configuration generation. SLM\_netconfig also incorporates an agent-based architecture that functions as an autonomous coordinator, interpreting user intents, validating outputs, and orchestrating the reasoning cycle to ensure syntactic and semantic correctness. This architecture enables adaptive, modular, and feedback-driven automation, allowing SLM\_netconfig to self-correct and efficiently handle complex configuration tasks.

SLM\_netconfig was compared to LLM-NetCFG. Results show that SLM\_netconfig achieves higher syntactic and semantic accuracy than our previous work while reducing translation latency from 5–7.5 minutes to 1–5 minutes. Moreover, SLM\_netconfig produces more concise and interpretable configurations than LLM-NetCFG. Results underscore the combination of fine-tuning, prompt engineering, and agentic orchestration; SLM\_netconfig is a promising, scalable, and efficient solution for autonomous network configuration.

The remainder of this paper is organized as follows. Section II provides the necessary background, and Section III reviews related work. Section IV presents the architecture and components of SLM\_netconfig, while Section V describes the fine-tuning process of small models for network configuration. Section VI outlines the dataset generation pipeline, and Section VII reports the experimental results. Finally, Section VIII concludes the paper and discusses future research directions.

\section{Background}
\subsection{Fine-Tuned and Small Language Models}
An LLM is a highly sophisticated architecture, typically comprising billions or even trillions of parameters, trained on extensive and diverse datasets for general-purpose language understanding and generation \cite{10614634}. LLMs constitute the foundation of numerous artificial intelligence applications but remain computationally expensive and resource-intensive to deploy \cite{Dur@Enh}. 

SLMs and FLMs have emerged to address the constraints of LLMs and meet the needs in particular application domains. SLMs are conceived with an emphasis on efficiency and specialization from their initial design and training stages, rendering them inherently faster and more resource-efficient, albeit with a narrower knowledge scope \cite{Jov@Com}. Such models are typically derived using techniques such as knowledge distillation—wherein a large teacher model transfers knowledge to a smaller student model \cite{hinton2015distillingknowledgeneuralnetwork}. Quantization reduces the numerical precision of model parameters, activations, or gradients to minimize computational overhead \cite{polino2018modelcompressiondistillationquantization}. Pruning eliminates redundant weights, neurons, or layers to decrease model size and inference cost \cite{Wang_2020}.

An FLM is a pre-trained model subsequently adapted to a specialized task through fine-tuning \cite{Ste@Fin, Han@Sys}. While fine-tuned LLMs remain computationally demanding, fine-tuned SLMs offer a highly efficient and compact alternative, maintaining strong performance on domain-specific tasks. The performance and resource requirements of any fine-tuned model depend on its underlying base model, with fine-tuning serving as a targeted layer of customization. Common fine-tuning strategies include transfer learning and instruction tuning \cite{Che@Fin, Nic@Int}. Transfer learning refines an entire pre-trained model using limited task-specific data, leveraging prior linguistic and semantic knowledge to enhance performance \cite{Col@Exp}. Instruction tuning, by contrast, trains models on instruction-formatted datasets (input–output pairs), thereby improving their ability to accurately interpret and follow user queries \cite{She@Ins}.

Several specialized techniques extend fine-tuning. Task-Adaptive Pre-training \cite{Gur@Don} involves pre-training on task-specific data before fine-tuning, enabling the model to capture task-related nuances. Domain-Adaptive Pre-training \cite{Gur@Don} focuses on domain-specific data, enhancing contextual understanding and vocabulary relevance. Alignment tuning ensures the model adheres to human values and expected behaviors, often via Reinforcement Learning with Human Feedback, where human evaluators guide model refinement \cite{Ouy@Tra}. Parameter-efficient fine-tuning approaches, such as Low-Rank Adaptation (LoRA) and Adapter Layers, provide significant computational advantages \cite{LoRA}. LoRA integrates low-rank matrices within model layers, enabling task-specific adaptation while updating only a fraction of the parameters \cite{Edw@Lor}. This selective modification drastically reduces memory and computational demands, making LoRA particularly suitable for resource-constrained environments \cite{Van@Tow}. Crucially, LoRA preserves the pre-trained model's foundational knowledge while efficiently incorporating domain-specific adjustments \cite{Fic@Bey}. Adapter Layers allow partial tuning through additional layers \cite{Naf@Ada}; however, LoRA's matrix-based approach offers even greater efficiency for high-performance, domain-targeted adaptation tasks, such as autonomous network configuration.

Prompting offers another parameter-efficient tuning technique, such as Priming, zero-shot, one-shot, and few-shot \cite{Liu@Pre}, chain-of-thought \cite{Jas@Cha}, and self-consistency \cite{Xue@Sel}, by keeping the model parameters frozen and learning continuous prompts that guide model behavior without modifying the core model. These techniques benefit task adaptation when computational resources are limited, providing lightweight customization without extensive parameter adjustment \cite{Les@The}. A prompt is a statement or question that guides an LLM output. It sets the context and specifies the task required by a system, an assistant, or a user role. In the system role, a well-designed prompt ensures the LLM accurately understands the user's request and generates a relevant response. The assistant role carries out the desired task based on the user's input, understands the user's request context, and tailors the response accordingly. The user role encompasses individual interactions with the LLM, such as entering queries, instructions, or commands to receive responses or perform tasks.

While the aforementioned techniques provide strategies for model adaptation, it is essential to recognize that most—particularly large-scale pre-training, extensive prompt optimization, and multi-stage transfer learning—entail substantial computational, temporal, and data resource demands. In contrast, fine-tuning, especially when applied to SLMs, achieves an optimal balance between efficiency and performance. More importantly, fine-tuning remains the only empirically validated method for mitigating hallucinations in specialized applications, such as network configuration generation. By adapting model parameters to domain-specific datasets, fine-tuning aligns the model's internal representations with the syntactic and semantic structures of the target domain, effectively reducing inconsistent or fabricated outputs and ensuring syntactic validity, contextual accuracy, and domain grounding.

\subsection{Agent Architectures}
An agent architecture facilitates the design of systems in which LLMs, SLMs, or FLMs operate as “agents,” capable of reasoning and decision-making through API invocations, code execution, or database queries, thus achieving predefined goals. Typically, an agent follows a perception–reasoning–action loop, wherein it interprets inputs, deliberates about the best course of action, and executes operations via external tools, iterating until task completion \cite{huang2024understanding}.

Beyond reactive text generation, agentic architectures enable the creation of autonomous, persistent, and goal-directed systems. Such agents integrate capabilities, including long-term memory, planning, self-reflection, and adaptive strategy development, allowing them to manage multi-step, dynamic, and context-sensitive tasks. These systems thus transcend traditional model–tool integrations, embodying self-directed, resilient behavior akin to that of artificial operators rather than passive executors \cite{xu2025mem}.

Within these architectures, fine-tuning and prompt engineering play complementary roles, jointly enhancing agent performance and adaptability. Fine-tuning embeds domain knowledge into the model, aligning its reasoning and decision-making processes with operational objectives. Prompt engineering functions as the behavioral interface that governs interaction flow, constrains reasoning trajectories, and modulates system behavior through structured role definitions and task instructions. The synergy between fine-tuning and prompt engineering enables the development of goal-oriented, low-hallucination agentic systems capable of executing complex, domain-specific tasks with precision and efficiency.

\section{Related Work}

Employing LLMs in network configuration is in its infancy; however, recent studies have demonstrated promising results. The NETBUDDY solution \cite{NETBUDDY} automates the translation of high-level natural language requirements into low-level network configurations, illustrating the capability of LLMs to balance accuracy, complexity, and computational cost in configuration-related tasks \cite{Fic@Aid}. Furthermore, this approach benchmarks the LLM's performance in generating network configurations for routing algorithms using two GPT-4-based prototypes.

GeNet \cite{GeNet} introduces a multimodal LLM framework that assists network engineers in designing and updating topologies by interpreting both visual and textual inputs, thereby reducing the need for manual intervention \cite{Fau@Ani}. The study in \cite{don2024llm} evaluates LLMs' comprehension of network structures, finding that while these models effectively handle basic configuration tasks, they struggle with complex topologies, highlighting prompt engineering as a potential mitigation strategy. \cite{Wha@Mon} integrates LLMs with formal verifiers to synthesize correct router configurations, enhancing reliability through localized feedback mechanisms. \cite{LLM@Mek} further demonstrates the applicability of LLMs in Intent-Driven Service Configuration for next-generation networks by translating user intents into Network Service Descriptors validated in real-world environments, effectively streamlining service deployment.

S-Witch \cite{Jeo@Swi} exemplifies the applicability of LLMs for automating the configuration of traditional network devices, thereby extending their relevance beyond software-defined paradigms. It introduces an intelligent assistant capable of generating command-line interface instructions for legacy switches based on natural-language intents, with validation performed in a network digital twin. Likewise, LLM-NetCFG \cite{Osc@Lar} leverages an LLM to automate configuration generation and integrates intent-based verification. By employing a local LLM, this approach mitigates the privacy and security concerns associated with Internet-based platforms, thereby preventing the exposure of sensitive configuration data.

LLNet~\cite{LLNet} introduces an intent-driven architecture that employs LLMs and SLMs to translate natural language intents into intermediate JSON representations, which are subsequently compiled into executable network programs for diverse data-plane (e.g., P4 and eBPF) and control-plane (e.g., Ryu and P4Runtime) environments. A key contribution of LLNet is the demonstration that SLMs can achieve over 88\% accuracy while substantially reducing computational overhead and energy consumption relative to larger models. However, LLNet primarily targets translation and deployment, and does not incorporate an automated, built-in verification mechanism to ensure that generated configurations conform to intended network policies before activation.

NetLLMBench~\cite{NetLLMBench} provides a benchmarking framework for systematically assessing LLM performance in network configuration tasks. It employs a closed-loop methodology that integrates prompt engineering with network emulation via Kathara, validating syntactic correctness through JSON schema checks and assessing semantic correctness in an emulated environment. The feedback loop facilitates iterative correction by the LLM, offering a structured evaluation process. Notably, NetLLMBench operates strictly as an evaluation platform rather than a configuration synthesis and assurance system; its primary focus is benchmarking tasks such as IP allocation and gateway selection rather than generating and formally verifying intent-driven configurations suited for production-grade deployments.

The NLI2Conf framework~\cite{NetworkCoPilot} integrates graph neural networks with an LLM (Llama 2) to update network configurations based on natural-language intents, while accounting for performance metrics such as delay and jitter. This framework represents multimodal network information including topology, configurations, and operational metrics and using a Text Attribute Graph. Trained on simulation-generated datasets, NLI2Conf advances intent-based configuration synthesis by embedding Quality of Service objectives into the generation process. However, its validation relies on post-hoc simulation with tools such as OMNet++, which cannot guarantee correctness or compliance with intent at scale. Moreover, simulation-based validation is computationally intensive and lacks the formal guarantees provided by static verification methods.

\begin{table*}[htbp]
\centering
\caption{Comparison of Related Works on LLM-Based Network Configuration Automation}
\label{tab:related_works_compact}
\renewcommand{\arraystretch}{1.3}
\setlength{\tabcolsep}{3pt}
\begin{tabular}{p{2.6cm} c c c c c c}
\hline
\textbf{Work} & \textbf{Intent} & \textbf{Config Gen.} & \textbf{Verification} & \textbf{Emulation} & \textbf{Hybrid} & \textbf{Closed Loop} \\
\hline
\cite{NETBUDDY} & \cmark & \cmark & -- & \cmark & -- & -- \\
\cite{GeNet} & \cmark & -- & -- & -- & -- & -- \\
\cite{Jeo@Swi} & \cmark & \cmark & -- & \cmark & -- & -- \\
\cite{Wha@Mon} & \cmark & \cmark & \cmark & -- & -- & -- \\
\cite{don2024llm} & \cmark & -- & -- & -- & -- & -- \\
\cite{LLM@Mek} & \cmark & \cmark & -- & \cmark & -- & -- \\
\cite{LLNet} & \cmark & \cmark & -- & \cmark & \cmark & -- \\
\cite{NetLLMBench} & \cmark & -- & -- & \cmark & -- & -- \\
\cite{NetworkCoPilot} & \cmark & \cmark & -- & \cmark & -- & -- \\
\rowcolor{gray!35}
\textbf{SLM\_netconfig} & \cmark & \cmark & \cmark & -- & \cmark & \cmark \\
\hline
\multicolumn{7}{l}{\textbf{Legend:} \cmark = Supported \quad -- = Not Supported or Not Addressed} \\
\end{tabular}
\end{table*}

Table \ref{tab:related_works_compact} compares existing research on LLM-based network configuration automation, emphasizing the functional coverage of each approach across key operational dimensions. The Intent column indicates whether the system supports natural language intent understanding to interpret configuration requirements. Verification indicates whether the proposed framework includes mechanisms to validate the correctness of generated configurations before deployment. Emulation represents the presence of an environment, or a virtual testbed, for validating or simulating the generated configurations. Hybrid captures the integration of multiple learning paradigms, such as combining fine-tuned LLMs with rule-based or traditional network automation tools, to enhance adaptability and performance. Closed Loop identifies whether the system includes a feedback mechanism for continuous learning or automated refinement based on performance outcomes. While the solutions cited advance the concept of autonomous network self-configuration by demonstrating how LLMs can automate configuration tasks, detect errors, and comprehend network topologies, they still face critical challenges. Chief among these are the reduction of error-prone outputs that can result in misconfigurations and operational inefficiencies \cite{Kuk@Pro}. SLM\_netconfig introduces a hybrid architecture that combines SLMs, FLMs, verification, and agents (closed-loop automation) to address these challenges, thereby offering a more comprehensive approach to autonomous network management.

\section{SLM\_netconfig}
This section outlines SLM\_netconfig and details the architecture, operation, and modules of our approach. 

\subsection{Architecture overview}
\begin{figure*}[!ht]
\centering
\caption{SLM\_netconfig Operation.}
\includegraphics[width=0.75\textwidth]{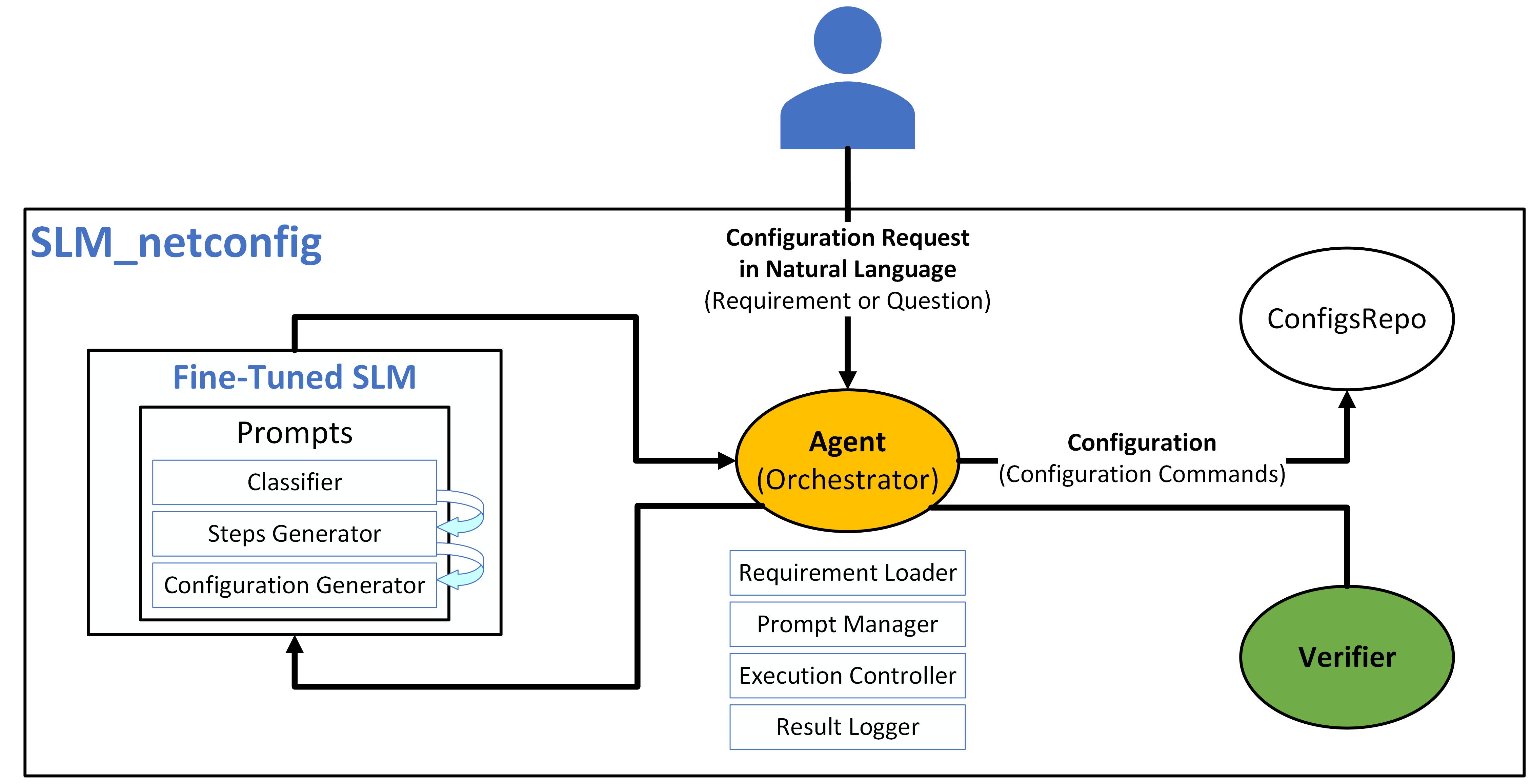}
\label{fig_architecture++}
\end{figure*}

Figure~\ref{fig_architecture++} illustrates the architecture of SLM\_netconfig. By combining FLMs and SLMs with agentic reasoning and formal verification, SLM\_netconfig bridges persistent scalability, reliability, and assurance gaps in LLM-based configuration systems, delivering a lightweight yet autonomous framework for safe, verifiable, and efficient network automation. The inclusion of an SLM component enables SLM\_netconfig to achieve high computational efficiency, rapid response times, and enhanced data privacy. Although SLMs possess less general knowledge than LLMs, this limitation does not hinder their effectiveness in network configuration tasks. Through domain-specific fine-tuning, the small model retains only knowledge relevant to networking, thereby improving accuracy and reasoning precision without compromising performance; this targeted specialization allows SLM\_netconfig to remain lightweight while delivering expert-level domain understanding centered on autonomous network self-configuration. SLM\_netconfig adopts an agent-based architecture that operates through a perception–reasoning–action cycle, coordinating modules for task management, verification, and self-reflection. The agent autonomously interprets user intents, synthesizes configurations, and iteratively verifies correctness through reasoning-driven refinement, introducing modularity, persistence, and adaptive intelligence.

The operational workflow of SLM\_netconfig proceeds as follows. \numrounded{1} The \textit{Agent} receives a natural-language configuration request, expressed as a requirement or question from a network practitioner, and employs the \textit{Classifier}, \textit{Steps Generator}, and \textit{Configuration Generator} prompts. \numrounded{2} The fine-tuned SLM\_netconfig model processes these prompts to classify intents, generate intermediate representations, and produce final actionable configurations, i.e., command sequences interpretable by network devices. \numrounded{3} The Verifier module examines each configuration to ensure correctness and applicability. This verification is essential, as configurations may span multiple devices or involve interdependencies requiring validation. \numrounded{4} If the Verifier detects an issue, it generates a feedback report with suggestions. The Agent uses this report to construct a refined prompt and re-invoke the fine-tuned question-configuration model to adjust the output accordingly.
\numrounded{5} This iterative cycle continues until the configuration is approved or a maximum iteration threshold is reached, preventing infinite refinement loops. \numrounded{6} Upon approval, the Verifier notifies the Agent, which transfers the verified configuration to the ConfigsRepo for deployment and reuse. This workflow enables SLM\_netconfig to interpret, generate, and validate network configurations with minimal human oversight. The following subsections describe the functionalities of the Agent, Verifier, and Model modules.

%The foundations of FSLM-NetCFG++ draw from the limitations observed in existing intent-based configuration systems such as NETBUDDY, GeNet, and S-Witch. While these frameworks illustrate the potential of LLMs to translate user intents into executable configurations, they exhibit deficiencies related to computational inefficiency, hallucination, and insufficient domain grounding. FSLM-NetCFG++ overcomes these challenges by employing parameter-efficient fine-tuning—specifically Low-Rank Adaptation (LoRA)—to embed configuration syntax and semantics directly within small, task-specialized models. This strategy improves syntactic precision, strengthens semantic alignment, and substantially reduces hallucination and resource consumption, enabling dependable performance in resource-constrained environments. 

\subsection{Modules}
\subsubsection{Agent} This module functions as the central orchestrator of SLM\_netconfig, managing task execution and data exchange between the Model and the Verifier to ensure the autonomous generation of accurate network configurations. It receives user-defined intents, structures them into formal prompts, and forwards these to the fine-tuned Model for translation into device-level configurations. Once a configuration is produced, the Agent invokes the Verifier for quality assurance and, when necessary, reinitiates refinement loops. This autonomous mechanism minimizes human intervention, enabling the system to handle complex configurations and enhance the efficiency and scalability of network configuration management.

The Agent includes a Requirement Loader, a Prompt Manager, an Execution Controller, and a Result Logger. The Requirement Loader retrieves network configuration intents for processing. The Prompt Manager dynamically formats and dispatches structured prompts for classification, goal decomposition, and configuration generation. The Execution Controller manages sequential task execution, tracks inference time, and ensures data integrity. The Result Logger aggregates, timestamps, and persistently stores all intermediate and final outputs for auditing and analysis. Collectively, these components enable the Agent to function autonomously, ensuring a reliable pipeline from intent intake to the generation of verified configurations.

\subsubsection{Verifier} SLM\_netconfig embeds a comprehensive verification mechanism directly into the configuration generation pipeline, an element largely absent in existing LLM-based network configuration frameworks. The Verifier ensures the syntactic, semantic, and operational validity of configurations generated by the fine-tuned model. Serving as a built-in quality assurance mechanism, it automatically inspects each configuration before deployment. Specifically, it checks for syntax compliance, adherence to device-specific standards, and alignment with the intended network objective. Beyond syntactic validation, this module performs semantic analysis to detect logical inconsistencies, dependency violations, and potential conflicts that could impact network performance. 

When errors are identified, the Verifier produces structured feedback, which the Agent uses to reframe prompts for iterative regeneration. This closed-loop validation mechanism increases configuration robustness, mitigates deployment risks, and enhances reliability, stability, and compliance in production environments.

\subsubsection{Model} At the core of SLM\_netconfig lies a fine-tuned small model that autonomously interprets and translates natural-language configuration intents into precise network configurations. The model receives configuration requests, either as requirements or questions, and generates the corresponding command sequences. Table~\ref{tab:que-config} exemplifies the model’s input (Question), intermediate output (Intermediate-Setup), and final configuration (Configuration). The intermediate representation enriches the input with relevant contextual information, such as device setup procedures, thereby facilitating more accurate and consistent configuration generation.

The model employs a structured prompting strategy to perform its translation tasks through three main prompt types: \textit{Requirement Classifier}, \textit{Steps Generator}, and \textit{Configuration Generator}. Each of these follows a standardized three-role structure that promotes clarity, modularity, and reproducibility across all prompt executions, enabling prompt types to be easily interpreted and debugged within the configuration generation pipeline. Each role (\textit{system}, \textit{assistant}, and \textit{user}) serves a distinct and complementary purpose. The \textit{system} role establishes the prompt's overall objective, providing the model with explicit task definitions and constraints. The \textit{assistant} role determines the structure and expected format of the output, ensuring that responses are consistent, machine-readable, and aligned with the task requirements. The \textit{user} role provides the contextual information, such as network requirements or the administrator's questions, that the model must analyze to perform the instructed task. 

\begin{itemize}
    \item The \textit{Requirement Classifier Prompt} instructs the model to identify the type of network configuration intent, such as admission control rules or device setups. The \textit{system} role guides the model in the classification task, the \textit{assistant} role defines the required output structure (e.g., specifying the intent type in the first line), and the \textit{user} role provides both the administrator’s intent and the set of possible intent categories. 
    \item The \textit{Steps Generator Prompt} is designed to produce a structured sequence of actions or sub-tasks that must be completed to fulfill the overall requirement goal. These steps do not represent configuration commands directly but instead outline the independent configurations, device preparations, or prerequisite setups required to achieve the intended outcome. By decomposing the high-level intent into modular steps, the prompt enables the model to reason over dependencies and ensure that all necessary components, such as interface preparation, authentication setup, or routing prerequisites, are logically and systematically covered before the configuration commands are generated. In this prompt, the \textit{system} role directs the translation process, the \textit{assistant} determines the output structure, and the \textit{user} inputs the intent along with contextual information such as network status or dependencies.
    \item The \textit{Configuration Generator Prompt} positions the model as a virtual network administrator tasked with generating vendor-specific configuration commands. In this context, the \textit{system} role defines the operational context, while the \textit{assistant} role specifies the desired configuration formats. To ensure automation compatibility, the model is explicitly instructed to output only the configuration commands without additional explanations and to separate configurations for different devices using the special delimiter “\texttt{\textasciitilde{}\textasciitilde{}\textasciitilde{}}”.
\end{itemize}

The fine-tuned small model interacts with the Agent to receive structured prompts and return formatted responses, and with the Verifier to validate the functional correctness and compliance of its outputs. This modular integration ensures that domain-specific knowledge, injected via fine-tuning, is fully leveraged during inference. Furthermore, the optimized single-model structure allows SLM\_netconfig to achieve high levels of accuracy, efficiency, and adaptability throughout the configuration process. The processes of creating and operating the fine-tuned small question-configuration model, as well as the datasets related to it, are presented in Sections \ref{fine-tuning} and \ref{pipeline}.

\section{Fine-tuning small language models for network self-configuration}
\label{fine-tuning}
\begin{figure}[h!]
\centering
\caption{Fine-tuning network configuration models}
\includegraphics[width=0.49\textwidth]{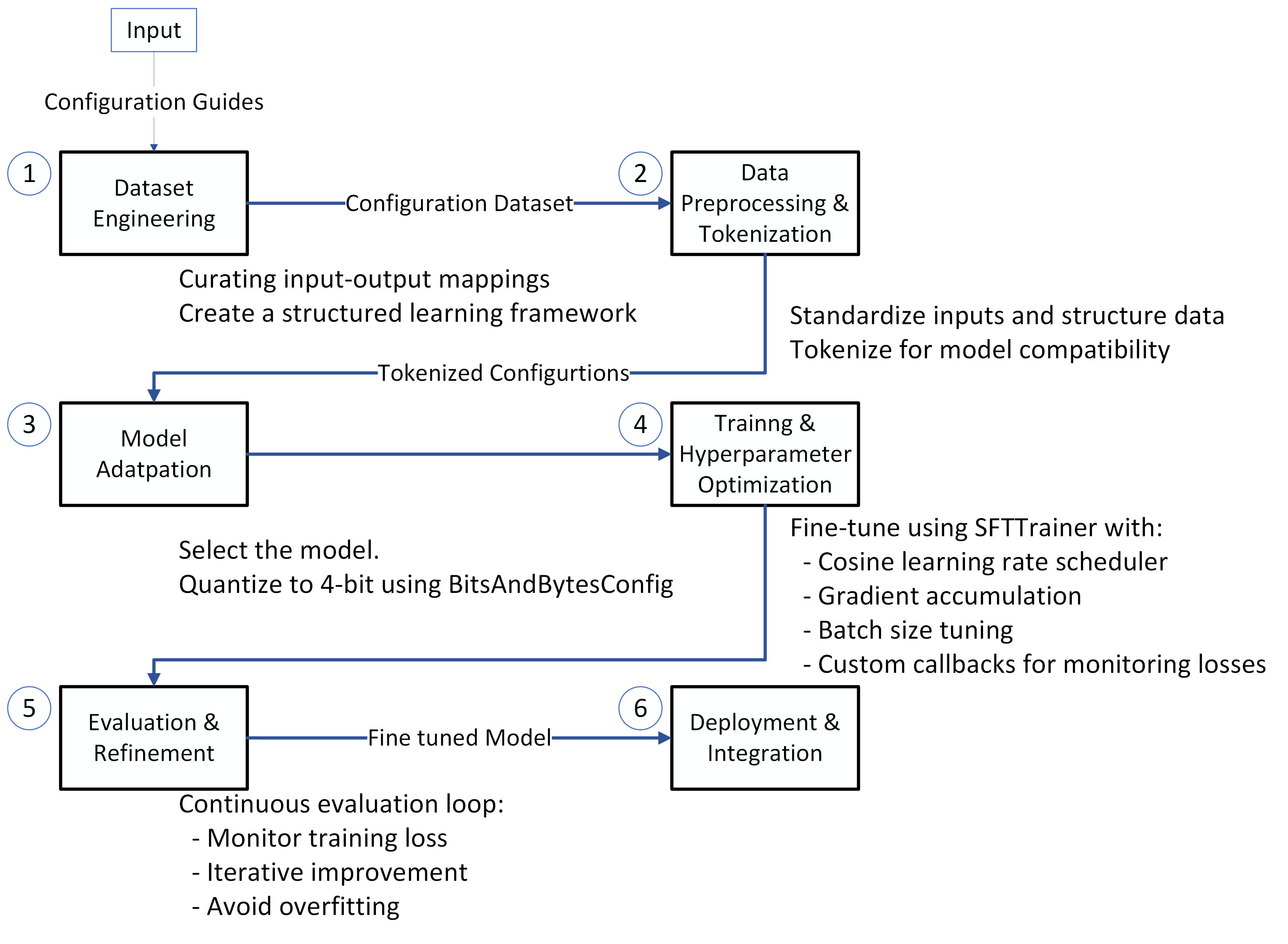}
\label{fig_Arq_conce}
\end{figure}

Figure~\ref{fig_Arq_conce} illustrates the fine-tuning pipeline of SLM\_netconfig, designed to transform a pre-trained SLM into a domain-optimized model that can autonomously perform network configuration tasks in self-configuration environments. The overall pipeline comprises six key stages: dataset engineering, data preprocessing, model adaptation, training and optimization, evaluation and refinement, and deployment and integration. Among these stages, dataset engineering stands out as a significant contribution of our work, as it establishes the foundation for effective fine-tuning. Through the proposed dataset-engineering pipeline, extensive, unstructured vendor documentation was systematically converted into high-quality, structured datasets that capture explicit mappings between intents (requirements or questions) and their corresponding configurations. This methodological advancement ensures that the fine-tuning process operates on coherent, domain-specific data, thereby enhancing model specialization, accuracy, and stability during training. Consequently, dataset engineering not only enables but also drives the effectiveness of the fine-tuning stage by providing the necessary semantic alignment and contextual richness for accurate intent-to-configuration translation.

The fine-tuning pipeline operates as follows. \numrounded{1} Dataset Engineering constructs domain-specific datasets that map natural-language intents to their corresponding device-level configurations via prompt–response pairs. The curated data, sourced from vendor documentation, expert manuals, and standardized configuration guides, ensures comprehensive coverage and high data quality. Tables \ref{tab:req-config} and \ref{tab:que-config} exemplify the mapping of requirements and questions to multi-step configurations, demonstrating how the model learns to associate intents with precise outputs. Proper dataset engineering is critical, as it directly influences the model’s capacity to generalize across real-world network scenarios.

\numrounded{2} Preprocessing and Tokenization standardize inputs, re-format prompt-response pairs, and structure the data to align with the model’s expected input distribution. Preprocessing mitigates noise and enhances training efficiency. Tokenization ensures that complex configuration commands and domain-specific terminology are accurately represented in the input space, enabling the model to capture both syntactic and semantic information relevant to network configuration tasks.

\numrounded{3} Model adaptation fine-tunes the base SLM through parameter-efficient optimization, specializing it for configuration generation. Instead of updating all parameters, only selected layers are adjusted, allowing the model to acquire domain-specific expertise while preserving general language competence. This selective updating approach significantly reduces computational overhead while maintaining the robustness of the pretrained model.

\numrounded{4} Training Dynamics are carefully driven by hyperparameter optimization. Variables such as learning rate, batch size, and gradient accumulation steps are fine-tuned to balance convergence speed, stability, and overall performance. Fine-tuning involves monitoring both short-term and long-term training behaviors to prevent oscillations or divergence, ensuring the model efficiently captures complex relationships between requirements and configurations.

\numrounded{5} Continuous Evaluation ensures that fine-tuning achieves its intended effect. Training loss and validation performance metrics guide refinements, preventing overfitting and ensuring that the model generalizes effectively to unseen configurations. Iterative improvement cycles, incorporating diverse test cases and edge scenarios, further enhance adaptability and precision, enabling the model to handle variations in vendor-specific configurations and network intents.

\numrounded{6} Integration into SLM\_netconfig: The fine-tuned SLM is incorporated into the SLM\_netconfig architecture to encapsulate the domain-specific expertise acquired during fine-tuning. The Orchestrator interacts directly with the SLM by sending natural language requirements as structured prompts and receiving configuration outputs. By embedding the fine-tuned SLM, the system seamlessly integrates into the configuration generation workflow, enabling automated processing of intents with high domain accuracy and contextual relevance. This integration ensures that the model’s specialized knowledge is fully leveraged in real-world network configuration scenarios.

In the following paragraphs, we illustrate how the fine-tuning pipeline can be employed to create specialized intent-centered translation models: one trained on requirement-configuration pairs and another on question-configuration pairs. The fine-tuned small requirement-configuration and question-configuration models translate network configuration natural-language intents, whether as requirements or queries into actionable device configurations, as shown in Tables \ref{tab:req-config} and \ref{tab:que-config}. Both models were developed following the same workflow presented in Figure \ref{fig_Arq_conce}.

\begin{table*}[ht]
    \centering
    \scriptsize
    \caption{Requirement-Configuration Example}
    \label{tab:req-config}
    \renewcommand{\arraystretch}{1.5}
    \begin{tabular}{|p{5cm}|p{10cm}|}
        \hline
        \textbf{Requirement} & \textbf{Intermediate-Setup} \\
        \hline
        Configure a port for Link Fault RFI Support by putting it into a blocking state when an OAM PDU control request packet is received with the Link Fault Status flag set. & 
        1. Enable \newline
        2. Configure terminal \newline
        3. Interface type number \newline
        4. \texttt{ethernet oam remote-failure \{critical-event | dying-gasp | link-fault\} action \{error-block-interface\}} \\
        \hline
        \multicolumn{2}{|c|}{\textbf{Configuration}} \\
        \hline
        \multicolumn{2}{|p{15cm}|}{
        \texttt{
        requirement: Configure a port for Link Fault RFI Support by putting it into a blocking state when an OAM PDU control request packet is received with the Link Fault Status flag set. \newline
        configuration: 1. Enable \newline
        2. Configure terminal \newline
        3. Interface type number \newline
        4. ethernet oam remote-failure \{critical-event | dying-gasp | link-fault\} action \{error-block-interface\}
        }} \\
        \hline
    \end{tabular}
\end{table*}

\begin{table*}[ht]
    \centering
    \scriptsize
    \caption{Question-Configuration Example}
    \label{tab:que-config}
    \renewcommand{\arraystretch}{1.5}
    \begin{tabular}{|p{5cm}|p{10cm}|}
        \hline
        \textbf{Question} & \textbf{Intermediate-Setup} \\
        \hline
        How can we ensure a network with a broader reach and an accurate representation of end-user experience while maintaining ease of deployment? & 
        To achieve a network with a broader reach and an accurate representation of the end-user experience, we can use Cisco devices ranging from low-end to high-end, including switches. Specifically, we can configure IP SLAs (Service-Level Agreements) on these devices to monitor application performance and ensure our network delivers the expected quality of service. \newline

        \texttt{Router\# configure terminal} \newline
        \texttt{Router(config)\# ip sla 1} \newline
        \texttt{Router(config-ip-sla)\# type http} \newline
        \texttt{Router(config-ip-sla-http)\# destination-ip <destination\_ip>} \newline
        \texttt{Router(config-ip-sla-http)\# source-ip <source\_ip>} \newline
        \texttt{Router(config-ip-sla-http)\# timeout 5000} \newline
        \texttt{Router(config-ip-sla-http)\# frequency 30} \\
        \hline
        \multicolumn{2}{|c|}{\textbf{Configuration}} \\
        \hline
        \multicolumn{2}{|p{15cm}|}{
        \texttt{
        Question: How can we ensure a network with a broader reach and accurate representation of end-user experience while maintaining ease of deployment? \newline
        Configuration: To achieve a network with broader reach and an accurate representation of the end-user experience, we can use Cisco devices ranging from low-end to high-end, including switches. Specifically, we can configure IP SLAs (Service-Level Agreements) on these devices to monitor application performance and ensure our network delivers the expected quality of service. \newline
        Router\# configure terminal \newline
        Router(config)\# ip sla 1 \newline
        Router(config-ip-sla)\# type http \newline
        Router(config-ip-sla-http)\# destination-ip <destination\_ip> \newline
        Router(config-ip-sla-http)\# source-ip <source\_ip> \newline
        Router(config-ip-sla-http)\# timeout 5000 \newline
        Router(config-ip-sla-http)\# frequency 30
        }} \\
        \hline
    \end{tabular}
\end{table*}

(1) Dataset engineering: To support specialization in network configuration generation, we constructed two structured datasets of intent-to-configuration. The first dataset, the requirement-configuration dataset, maps explicit configuration requirements extracted from operational and instructional documents to the correct device configuration outputs. The second dataset, the question-configuration dataset, refines and extends the initial dataset by rephrasing configuration intents into question form and improving the quality and completeness of configurations. This dual-dataset design enables the system to comprehend different communication styles as direct intent expression and natural query-driven interaction, resulting in a more versatile and user-aligned model. The process for creating the requirement-configuration dataset is outlined in Section V, while Section VI.C provides the methodology for developing the question-configuration dataset.

(2) Data preprocessing and tokenization: To prepare the datasets for fine-tuning, we standardized them into prompt-response training samples using two consistent templates: \texttt{requirement: <requirement>\textbackslash answer: <answer>}, and \texttt{question: <question>\textbackslash answer: <answer>}. We employed the tokenizer of the selected SLM and introduced [BOS] and [EOS] tokens to mark the start and end of each sequence, thus enabling compatibility with causal language modeling. Right-side padding was applied using the EOS token to maintain predictable alignment during batch processing and to minimize tokenization artifacts. This uniform formatting reduced noise during learning and ensured that the input distribution matched the base model's expectations, improving convergence and output consistency.

\begin{figure}[h!]
\centering
\caption{Syntax Accuracy of SLM\_netconfig models - Question vs Requirement}
\includegraphics[width=0.48\textwidth]{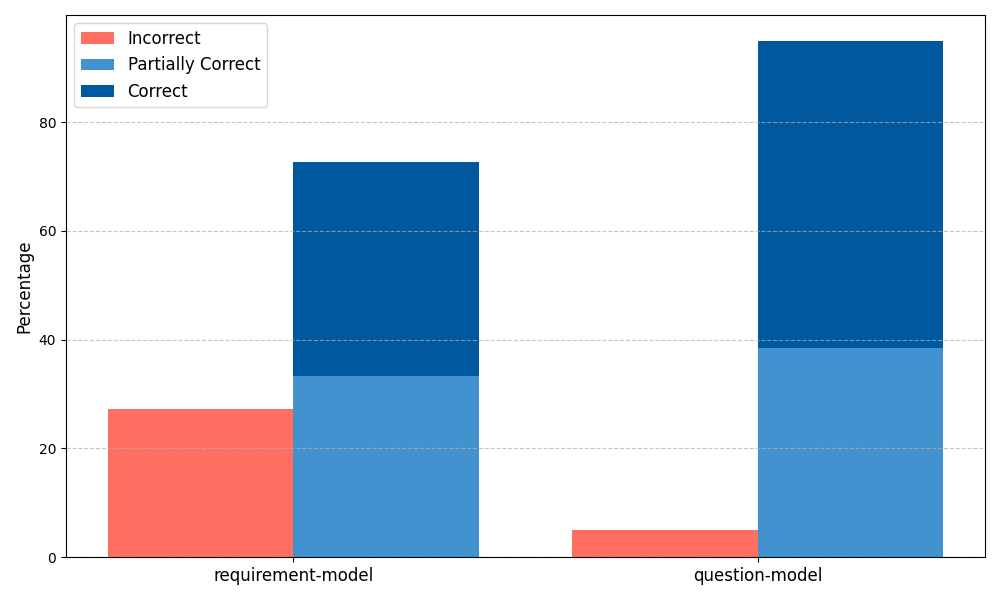}
\label{fig:syntax3}
\end{figure}

\begin{figure}[h!]
\centering
\caption{Format Accuracy of SLM\_netconfig - Question vs Requirement}
\includegraphics[width=0.48\textwidth]{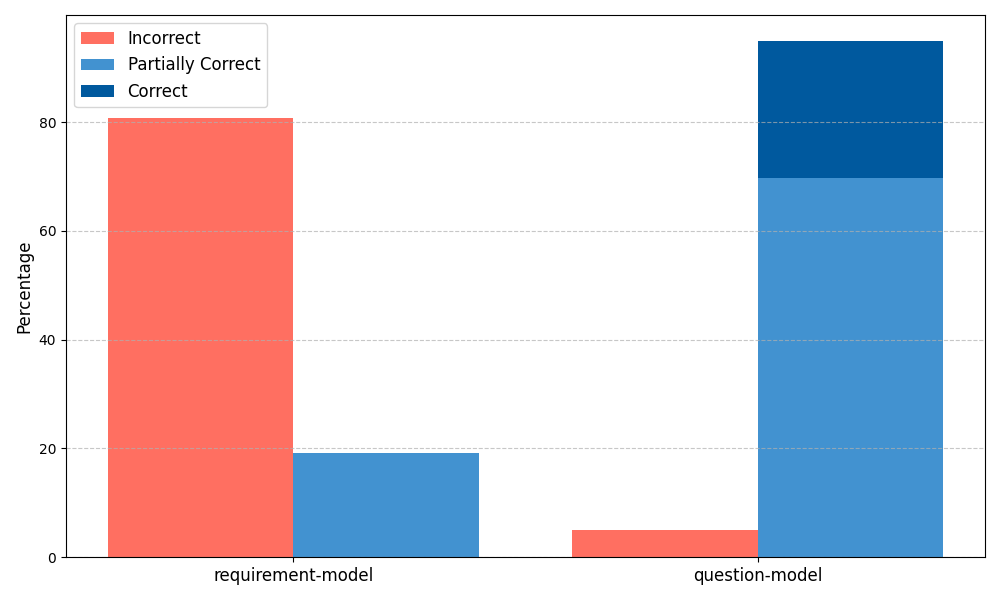}
\label{fig:format3}
\end{figure}

\begin{figure}[h!]
\centering
\caption{Goal Accuracy of SLM\_netconfig - Question vs Requirement}
\includegraphics[width=0.48\textwidth]{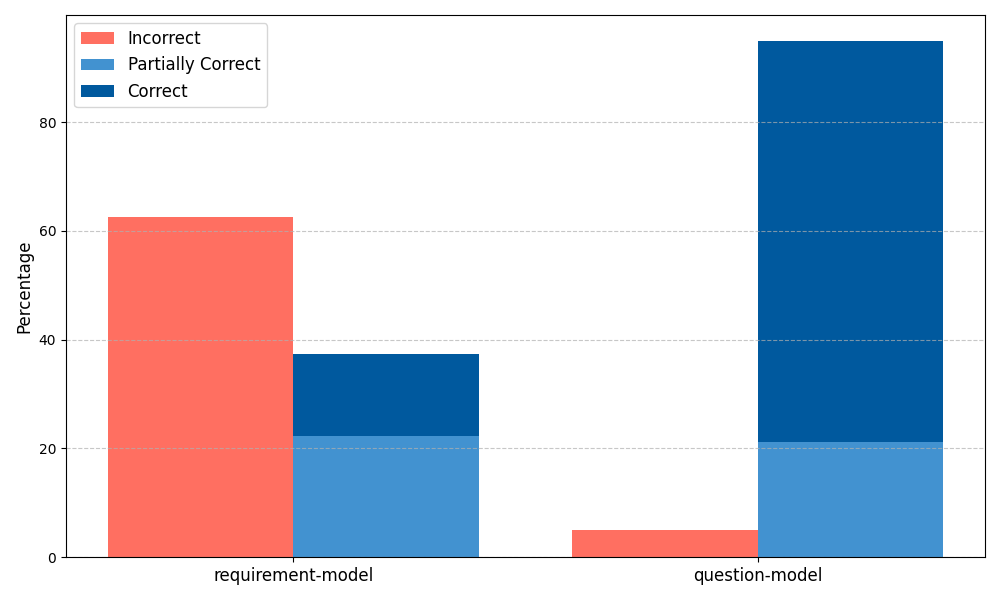}
\label{fig:configuration3}
\end{figure}

(3) Model adaptation: Zephyr-7B was chosen as the foundational SLM for fine-tuning because it offers an excellent balance between performance and computational efficiency compared to larger LLMs. To optimize GPU memory usage, we applied 4-bit quantization via BitsAndBytesConfig, which significantly reduces the model's footprint while preserving representational quality. Additionally, LoRA was adopted to fine-tune only a small subset of trainable parameters, enabling the model to specialize in network configuration tasks without degrading its broader language understanding. Importantly, two distinct models were produced: one trained on the requirement-configuration dataset and the other trained on the refined question-configuration dataset. Experimental results illustrated in Figures~\ref{fig:syntax3}, \ref{fig:format3}, and \ref{fig:configuration3}
show that the question-configuration model outperforms the requirement-configuration model across syntax correctness, formatting accuracy, and configuration quality—even when handling requirement- and question-style inputs. Consequently, this superior model is designated as the final SLM\_netconfig configuration generator.

(4) Training and hyperparameter optimization: Fine-tuning was conducted using Hugging Face’s SFTTrainer with a cosine learning rate scheduler to facilitate smooth parameter adjustments throughout the training process. An optimized batch size and gradient accumulation strategy ensure stable learning, even with limited hardware resources. Additionally, we employed label-to-token alignment to reduce prediction variability. Early-stage training heuristics and checkpointing enabled iterative monitoring and rollback in the event of overfitting, ensuring that the final models maintained generalization while achieving task-specific performance improvements.

(5) Evaluation and refinement: Performance was continuously monitored through custom callbacks measuring training and validation loss, facilitating real-time corrective tuning of learning dynamics. Dataset resampling and error-pattern inspection mitigated bias and improved robustness across distinct types of configuration intents. The feedback loop between evaluation metrics and hyperparameter updates progressively enhances the syntactic correctness and semantic alignment of the generated configurations.

(6) Deployment: Upon achieving stable performance, the fine-tuned model question-configuration was deployed into the SLM\_netconfig  pipeline. This integration enables seamless natural-language intent interpretation and configuration generation during inference, supporting both question-style and requirement-style inputs. The deployed model maintains efficiency, low memory usage, and rapid execution, enabling privacy-preserving deployment in localized network management environments.

\section{Pipeline for building network configuration datasets}
\label{pipeline}
This section introduces the proposed pipeline for constructing network configuration datasets, along with two resulting datasets—requirement–configuration and question–configuration—generated through its application. The proposed pipeline enables the systematic transformation of extensive, unstructured network configuration guides into structured datasets suitable for training LLMs and SLMs.

\subsection{Overview}
\begin{figure*}[!ht]
\centering
\caption{Pipeline for creating network configuration datasets.}
\includegraphics[width=0.95\textwidth]{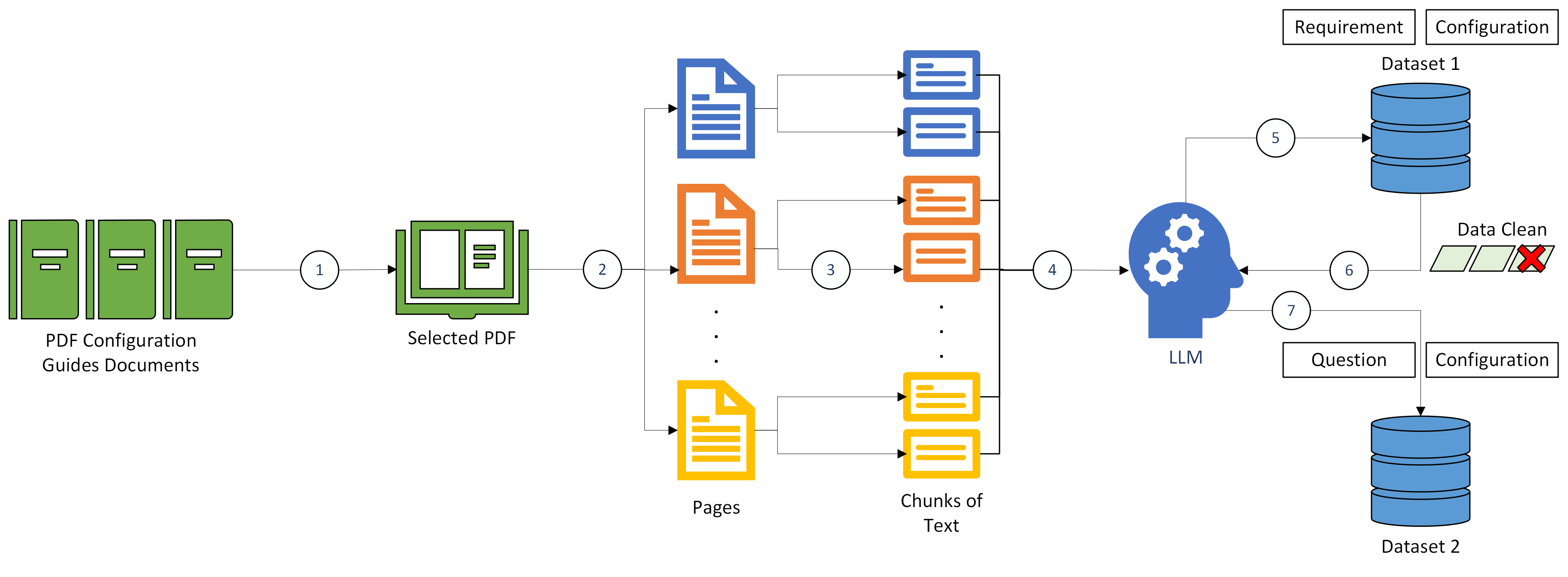}
\label{fig_DatasetCreate}
\end{figure*}

Figure~\ref{fig_DatasetCreate} depicts the proposed multi-stage pipeline for developing network configuration datasets. The pipeline comprises seven stages: \numrounded{1} Data Collection, \numrounded{2} Data Extraction, \numrounded{3} Page Handling, \numrounded{4} Chunking Data, \numrounded{5} Data Enhancement, \numrounded{6} Data Cleaning, and \numrounded{7} Data Refinement.
The Data Collection stage aggregates relevant information from diverse sources to ensure comprehensive domain coverage. Data Extraction isolates meaningful content from raw documents while preserving their logical and structural integrity. Page Handling organizes the extracted data into structured units, maintaining continuity across document sections. Chunking Data divides the text into manageable and semantically coherent segments, facilitating efficient processing by language models. Data Enhancement improves the clarity, consistency, and usability of extracted content. Data Cleaning eliminates redundancies, inconsistencies, and irrelevant elements to enhance data reliability. Data Refinement fine-tunes the resulting dataset to ensure semantic alignment and readiness for training. This pipeline was applied to generate the requirement–configuration and question–configuration datasets.

A key contribution of our work lies in the design and implementation of a fully automated, multi-stage pipeline for network configuration dataset engineering. This process bridges the gap between unstructured network documentation and structured data suitable for training language models. Unlike traditional approaches that rely heavily on manual data annotation or rule-based extraction, our pipeline introduces a scalable, semi-autonomous framework that transforms large volumes of heterogeneous technical content (e.g., vendor configuration guides) into coherent, machine-interpretable datasets. The proposed dataset engineering strategy differs fundamentally from existing dataset construction methodologies, which typically focus on high-level intent detection or natural language processing of network-related text. In contrast, our pipeline integrates LLM-driven semantic parsing (via Llama-3-8B) into the data enhancement phase to automatically extract and align “requirement–configuration” pairs.

%The proposed pipeline enables a fully automated, multi-stage process for engineering network configuration datasets, bridging the gap between unstructured vendor documentation and structured data suitable for training language models. Its novelty lies in the autonomous extraction, structuring, and refinement of requirement–configuration knowledge from heterogeneous, large-scale sources without manual annotation or handcrafted rules. Unlike existing methods that focus on intent classification or manually mapping textual descriptions to configurations, the proposed pipeline integrates LLM-driven semantic parsing directly into the data enhancement phase, enabling automatic alignment between configuration intents and their corresponding command sequences.

\subsection{Requirement-configuration dataset}
The requirement–configuration dataset serves as a foundational component for training SLM\_netconfig, enabling the model to learn mappings between high-level network requirements (intents) and corresponding configuration commands. This capability allows for the model to produce accurate, contextually appropriate configurations when processing new requests. The dataset was constructed following the stages defined in the proposed pipeline.

\subsubsection{Data Collection}
Relevant source materials were identified and collected to support the creation of an autonomous configuration agent. Specifically, official Cisco configuration guides corresponding to the Cisco IOS XE Gibraltar 16.11.1 release were selected from Cisco’s documentation portal. These guides, provided in PDF format, contain comprehensive configuration instructions, command sequences, and detailed explanations of the underlying operational requirements—making them ideal for developing data-driven configuration models.

\subsubsection{Data Extraction}
A Python-based extraction script was developed to retrieve textual content from the selected configuration guides. The script preserves document structure, including section hierarchies, lists, and command blocks, ensuring that contextual relationships within the data are maintained.

\subsubsection{Page Handling}
The extracted text was processed on a per-page basis using a custom Python script. The script output is a list of strings, each string representing the text extracted from a single page. Given the size and complexity of the guides, the extracted strings were unstructured and contained a mix of technical explanations, configuration commands, and networking concepts.

\subsubsection{Chunking Data}
We divided the per-page extracted data (i.e., an entry in the previously created list) via a Python script into manageable chunks to ensure that LLMs (or SLMs/FLMs) can process the text efficiently and accurately, minimizing the risk of incomplete or erroneous outputs. Each chunk included text segments of no more than 1000 characters; the segments were obtained by splitting the text at natural points, such as spaces or sentence boundaries, to avoid disrupting the flow of information and ensure that each chunk remained meaningful and contextually coherent.

\subsubsection{Data Enhancement}
We improved the quality of the chunked data in this step to ensure that LLMs (or SLMs/FLMs) can generate network configurations effectively. Each chunk of text was processed using the Llama-3-8B SLM \cite{Llama3} via a prompt carefully designed to identify and separate the "requirement" from the "configuration". For example,  the requirement might be "Enable VLAN routing on the router," while the configuration could include the specific commands necessary to accomplish this. 
We selected the Llama-3-8B model for this data processing task because of its ability to efficiently handle large volumes of data while maintaining high performance in parsing and structuring complex textual information. This capability enabled the quick and accurate processing of detailed technical network configuration guides, resulting in precise outputs. Moreover, using this SLM-based data processor minimizes the need for manual extraction and annotation of requirement-configuration pairs.

\subsubsection{Data Cleaning}
The enhancement process yielded a total of 5,505 requirement–configuration pairs. A custom Python script was subsequently employed to remove 401 incomplete or invalid pairs resulting from textual irregularities in the original documents, such as placeholders (e.g., “None specified in this text” or “N/A”).
By filtering out such pairs, the overall quality of the requirement-configuration dataset was significantly improved. The requirement-configuration dataset is available on github\footnote{\href{https://github.com/oscarGLira/Fine-tuned-Configuration-Agent.git}{https://github.com/oscarGLira/Fine-tuned-Configuration-Agent.git}}, including 5,104 pairs. An example of an entry in this dataset is as follows.

\begin{itemize}
    \item Requirement: ´´Enable OSPF routing on all interfaces."
    \item Configuration command: ´´router ospf 1 and network 192.168.1.0 0.0.0.255 area 0".
\end{itemize}

\subsection{Question-Configuration Dataset}
The question-configuration dataset enables SLM\_netconfig to effectively address realistic scenarios in which network administrators create specific configuration queries. We generated this dataset from the previous one by refining the data (as outlined in Step \numrounded{7} of Figure \ref{fig_DatasetCreate}) using a data enhancer. This enhancer rephrased the requirements as questions, enabling SLM\_netconfig to learn how to respond to network configuration requests expressed in an interrogative format. However, it is pivotal to emphasize that although the model was trained on data structured as questions, the underlying model (LLM/SLM) retains the capacity to comprehend intents expressed in non-question formats, such as declarative or imperative statements. This contextual understanding capability, characteristic of language models, ensures that the model’s performance is not restricted to the syntactic structure of the training data. Instead, the question-oriented format serves merely as a training strategy to enhance precision and consistency, while the model’s broader language comprehension enables it to interpret diverse expressions of intent naturally. Furthermore, it is noteworthy that the dataset standardizes terminology and ensures coherence between each intent and its corresponding configuration output. The question-configuration dataset is also available in CSV format on our GitHub, comprising 5097 pairs. An example entry is shown below:

\begin{itemize}
    \item Question: "How do I enable OSPF routing on all interfaces?".
    \item Configuration command: "router ospf 1 and network 192.168.1.0 0.0.0.255 area 0".
\end{itemize}

\section{Evaluation}
This section presents the evaluation of SLM\_netconfig. Subsection \ref{subsec:test_env_prot} depicts the test environment and the prototype used for fine-tuning and inference. Subsection \ref{subsec:metrics} presents the accuracy, time, and complexity metrics, while Subsection \ref{subsec:results} discusses and presents the results. Subsection \ref{highlights} summarizes the main findings of the evaluation.

\subsection{Environment and Prototype}
\label{subsec:test_env_prot}
\begin{figure}[!ht]
\centering
\caption{Test Environment.}
\includegraphics[width=0.4\textwidth]{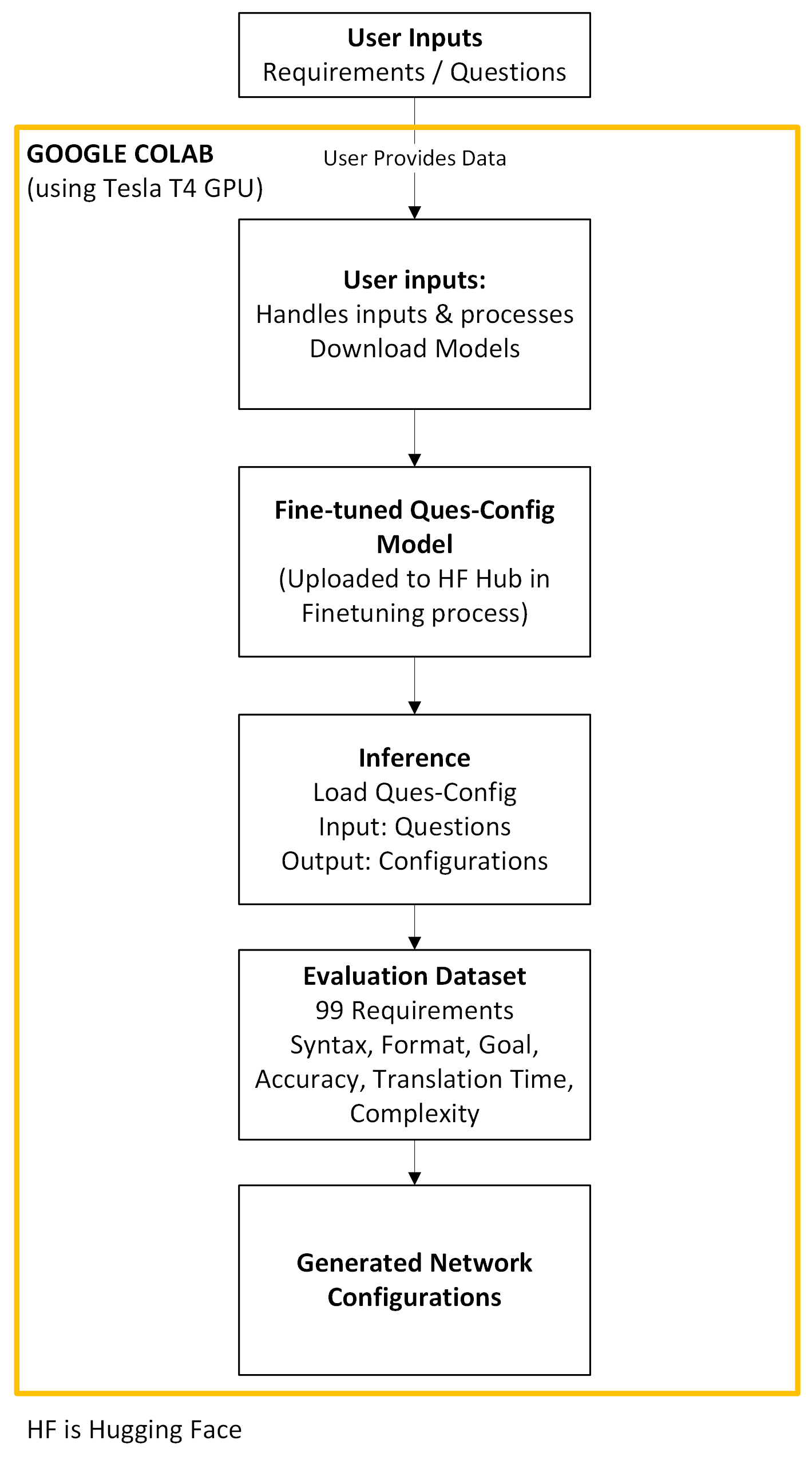}
\label{fig_Environment}
\end{figure}

Figure~\ref{fig_Environment} illustrates the prototype of SLM\_netconfig, which incorporates the Zephyr 7B fine-tuned small question-configuration model. We developed and tested SLM\_netconfig on Google Colab, which provides a flexible, accessible environment for ML tasks and seamless integration with powerful hardware. Specifically, a free Tesla T4 GPU was utilized for the processes mentioned above. We used Python as the primary programming language.

The development and deployment process of the SLM\_netconfig prototype proceeded as follows. First, a Python script interfaced with Hugging Face to retrieve the base pre-trained SML, Zephyr-7B-$\beta$. After downloading the base model, a dedicated Python script executed the fine-tuning procedure, updating the model parameters using the question-configuration dataset described earlier. Upon completion of fine-tuning, the resulting small model was uploaded to Hugging Face to enable convenient access, sharing, and subsequent inference.

Python scripts were also developed to orchestrate the entire inference pipeline. These scripts load the fine-tuned model from Hugging Face, process input data (e.g., configuration intents or natural-language questions), and generate the corresponding network configurations. The accuracy of the fine-tuned model was assessed for syntax and configuration-goal correctness using a dataset of 99 network configuration requirements. These requirements encompass a diverse range of network configuration domains, including interface initialization, IP addressing, access control lists, routing policies, and tunneling mechanisms. A separate dataset of 90 requirements was used to evaluate SLM\_netconfig during inference, assessing translation time and configuration complexity.

\subsection{Metrics}
\label{subsec:metrics}
We used various evaluation metrics to assess the performance of SLM\_netconfig, including syntax, format, configuration goals, translation time, and composite complexity. These metrics were applied to the fine-tuned small question-configuration model and LLM-NetCFG. The syntax accuracy, defined in Equations~\ref{eq:valid} and~\ref{eq:syn}, evaluates whether each generated configuration line $l_i$ adheres to valid Cisco IOS grammar. As shown in Equation~\ref{eq:valid}, the function $\text{valid}(l_i)$ returns $1$ if the line matches known IOS grammar, $0$ if it is incomplete or unusual, and $-1$ if it is grammatically invalid. The global syntax score (\texttt{SYN}) for the configuration is computed as shown in Equation~\ref{eq:syn}, where: $1$ means all lines in the generated configuration are valid, $0$ means there are incomplete or unusual responses, and $-1$ implies that at least one line in the configuration is invalid.

\footnotesize
\begin{equation}
\begin{split}
\text{valid}(l_i) = \begin{cases}
1 & \text{if } l_i \text{ matches known IOS grammar} \\
0 & \text{if } l_i \text{ is incomplete or unusual} \\
-1 & \text{if } l_i \text{ is invalid}
\end{cases}
\end{split}
\label{eq:valid}
\end{equation}

\begin{equation}
\begin{split}
\text{Syntax Accuracy} = \begin{cases}
1 & \text{if } \forall l_i,\ \text{valid}(l_i) = 1 \\
0 & \text{if } \exists l_i:\ \text{valid}(l_i) = 0,\\
-1 & \text{if } \exists l_i:\ \text{valid}(l_i) = -1
\end{cases}
\end{split}
\label{eq:syn}
\end{equation}
\normalsize

%The format accuracy, given by Equations~\ref{eq:id} and~\ref{eq:fmt}, evaluates whether the configuration output follows a template. Equation~\ref{eq:id} computes $\text{iD}(l_1)$ as a classification of the first line of the configuration: $1$ if it matches a template (e.g., device name with delimiters, like \texttt{\~{}\~{}\~{}Device1\~{}\~{}\~{})}), $0$ if it is just a plain device name (e.g., when the device name appears without the any delimiters: \texttt{Device1}), $-1$ if in other case. Equation~\ref{eq:fmt} calculates \texttt{FMT}, the overall format correctness, based on whether the rest of the configuration ($l_{i>1}$) follows expected patterns.

%\footnotesize
%\begin{equation}
%\begin{split}
%\text{iD}(l_1) = \begin{cases}
%1 & \text{if } l_1 \text{ matches } \sim\sim\sim\text{Device Name}\sim\sim\sim \\
%0 & \text{if } l_1 \text{ is a plain device name line} \\
%-1 & \text{otherwise}
%\end{cases}
%\end{split}
%\label{eq:id}
%\end{equation}

%\begin{equation}
%\begin{split}
%\text{Format Accuracy} = \begin{cases}
%1 & \text{if } \text{iD}(l_1) = 1 \text{ and all } l_{i>1} \text{ are configurations} \\
%0 & \text{if } \text{iD}(l_1) = 0 \text{ and all } l_{i>1} \text{ are configurations} \\
%-1 & \text{otherwise}
%\end{cases}
%\end{split}
%\label{eq:fmt}
%\end{equation}
%\normalsize

The goal configuration accuracy, defined by Equation~\ref{eq:cmd}, quantifies the extent to which a generated configuration $C$ satisfies the intended requirement $R$. A score of $1$ indicates full compliance with the specified  configuration requirement, 
%$0$ indicates partial coverage of key goals (the generated commands are relevant but incomplete or not fully functional,e.g., missing one required command)
a score of (0) denotes partial fulfillment, in which relevant but incomplete commands are produced (e.g., the omission of at least one mandatory command) , and $-1$ reflects a failure to satisfy the requirement.
%indicates that the configuration does not fulfill the requirement at all.

\footnotesize
\begin{equation}
\begin{split}
\text{Goal Accuracy} = \begin{cases}
1 & \text{if } C \text{ fully satisfies } R \\
0 & \text{if } C \text{ partially satisfies } R \\
-1 & \text{if } C \text{ does not satisfy } R
\end{cases}
\end{split}
\label{eq:cmd}
\end{equation}
\normalsize

The translation time metric, formulated in Equations~\ref{eq:total_len} to~\ref{eq:norm_time},
 measures the computational duration required for the evaluated models to translate a human-readable configuration requirement into a device-actionable configuration. Equation~\ref{eq:total_len} defines the \texttt{total\_len} as the sum of the lengths of the translated requirement and the generated configuration. This value is normalized to the interval [0, 1] via min-max scaling, as shown in Equation~\ref{eq:norm_len}, yielding  the \texttt{norm\_total\_len}. Equation~\ref{eq:duration} calculates the \texttt{total\_dura\_time} in minutes by summing the translation and configuration durations (in seconds) and dividing by 60. This duration is further normalized in Equation~\ref{eq:norm_time}.

\footnotesize
\begin{equation}
\text{total\_len} = \text{trans\_len} + \text{config\_len} 
\label{eq:total_len}
\end{equation}

\begin{equation}
\text{norm\_total\_len} =
\frac{ \text{total\_len} - \min(\text{total\_len}) }
     { \max(\text{total\_len}) - \min(\text{total\_len}) }
\label{eq:norm_len}
\end{equation}

\begin{equation}
\text{total\_dura\_time} =
\frac{\text{Translation Time} + \text{Configuration Time}}{60}
\label{eq:duration}
\end{equation}

\begin{equation}
\text{norm\_total\_dura\_time} =
\frac{ \text{total\_dura\_time} - \min(\text{total\_dura\_time}) }
     { \max(\text{total\_dura\_time}) - \min(\text{total\_dura\_time}) } 
\label{eq:norm_time}
\end{equation}
\normalsize

The complexity score provides a unified measure of textual and temporal configuration complexity. Equation~\ref{eq:composite_score} defines such a score in terms of the two normalized components: \texttt{norm\_total\_len} and \texttt{norm\_total\_dura\_time}.

\footnotesize
\begin{equation}
\text{complexity\_score} =
\frac{ \text{norm\_total\_len} + \text{norm\_total\_dura\_time} }{2} 
\label{eq:composite_score}
\end{equation}
\normalsize

%%%%% STOPPED here

\subsection{Results and Analysis}
\label{subsec:results}
\begin{figure}[!ht] 
\centering \caption{Syntactic Accuracy - LLM-NetCFG vs SLM\_netconfig.} 
\includegraphics[width=0.49\textwidth]{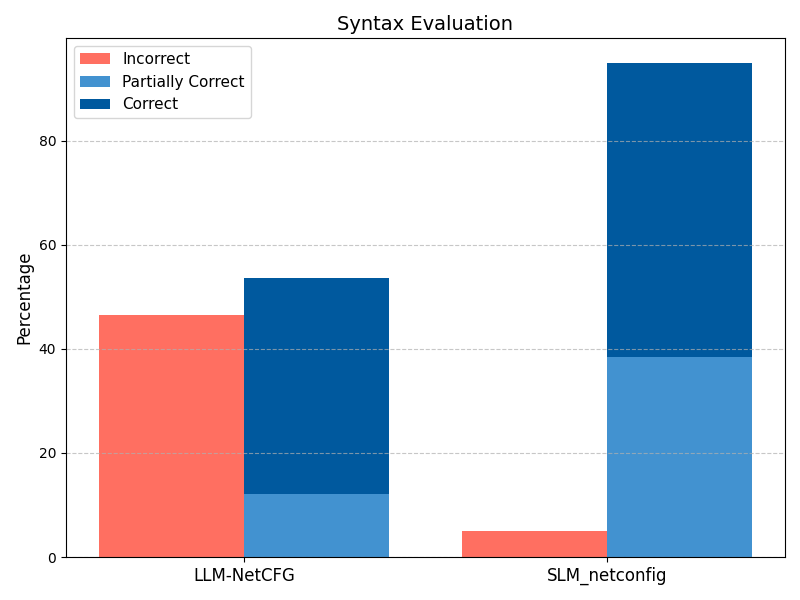} 
\label{fig_syntax} 
\end{figure}

Figure~\ref{fig_syntax} presents the results obtained by LLM-NetCFG and SLM\_netconfig (operating with the fine-tuned small question-configuration model) regarding the syntax metric when processing the network configuration requirements dataset with 99 samples. SLM\_netconfig achieved a higher syntactic accuracy (57\% correct and 38\% partially correct answers) than LLM-NetCFG (42\% correct and 12\% partially correct answers). The superior performance of the question-configuration model stems from the well-organized and refined input data used during fine-tuning, which enabled effective learning and reinforced syntactic consistency. Conversely, the lower syntactic accuracy of LLM-NetCFG highlights the need for domain-specific fine-tuning to improve precision when generating complex network configurations. These findings underscore the importance of structured and curated training data for improving syntactic correctness. Furthermore, they demonstrate that SLM\_netconfig is a promising solution for scenarios such as automated network provisioning and policy-driven configuration enforcement, where accurate and error-free configuration commands are operationally critical.

In addition to the quantitative improvements in syntactic accuracy, the partially correct category offers crucial insights into the behavior of both models; it includes configurations with the correct elements and parameters but that differ from the strict Cisco IOS syntax constraints required for command execution. For example, the models may apply slight variations in keyword usage or ordering that preserve functional meaning but remain incompatible with device parsers. While such cases still trigger errors when deployed directly, they contain recoverable information that can be corrected with lightweight automated post-processing. Thus, distinguishing partially correct from incorrect outputs reveals where the model demonstrates underlying comprehension despite failing to meet rigid CLI constraints.

\begin{figure}[h!]
\centering
\caption{Configuration Goal Accuracy - LLM-NetCFG vs SLM\_netconfig.}
\includegraphics[width=0.49\textwidth]{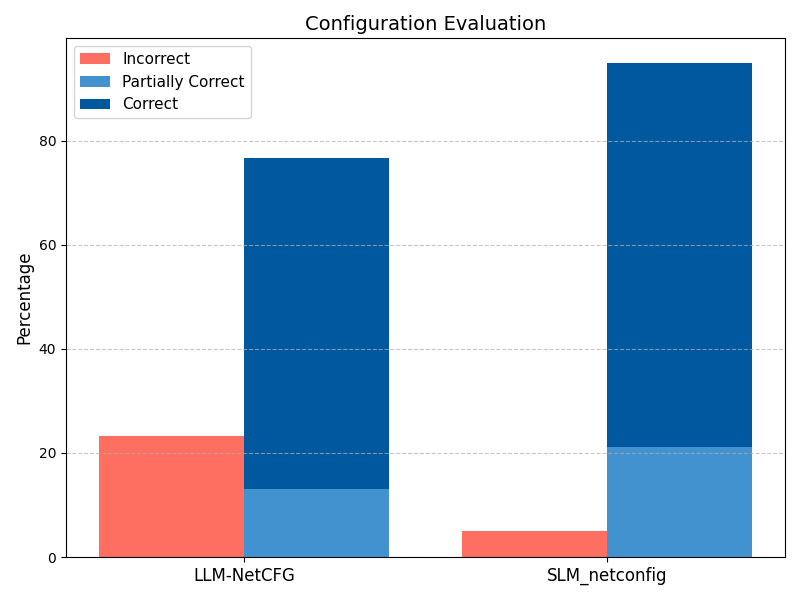}
\label{fig_Config}
\end{figure}

Figure~\ref{fig_Config} reports the configuration-goal accuracy for both models. LLM-NetCFG vs SLM\_netconfig (equipped with a fine-tuned question-configuration model) achieved the highest accuracy (74\%) on this metric, benefiting from the structured question-based dataset that provided explicit guidance during training, enabling the model to align generated configurations with target goals better. LLM-NetCFG achieved acceptable accuracy (64\%), reflecting the benefits of its broad pretraining across diverse textual sources, which enables it to generalize network configuration patterns even without domain-specific fine-tuning. Overall, these findings indicate that LLM-NetCFG vs SLM\_netconfig achieves superior goal-oriented configuration accuracy, making it highly promising for tasks such as network optimization and security policy enforcement, as well as other accuracy-critical automation tasks.

The partially correct cases in the goal metric represent an intermediate level of functional correctness, where the model captures the configuration intent but performs only a subset of the necessary steps. While these outputs may enable partial network operation, they risk performance or security gaps if deployed without additional adjustments. Differentiating these cases provides a more accurate depiction of model capability by avoiding the loss of meaningful information that would occur if partial successes were categorized as entirely incorrect, particularly in the automation of complex configuration tasks.

To further reduce partially correct outcomes across all evaluation dimensions, multiple enhancement strategies can be adopted. Prompt engineering offers improvements without modifying the model by enforcing template compliance, validating syntax through self-verification prompts, or explicitly enumerating required configuration steps before command generation. Fine-tuning remains the most effective strategy, as demonstrated by the improved performance of SLM\_netconfig, which enables internalization of critical command patterns and formatting structures. Retrieval-augmented generation (RAG) offers a viable approach, enabling dynamic access to authoritative configuration documentation or trusted repositories of valid command examples during inference. Moreover, allowing the model to access a dataset of accepted formatting examples can further improve the performance in formatting configuration commands by reinforcing consistent structural patterns during generation. Together, these strategies enhance the determinism, reliability, and operational suitability of configuration-generation systems.

\begin{figure}[h!]
\centering
\caption{Complexity per Requirement - LLM-NetCFG vs SLM\_netconfig.}
\includegraphics[width=0.49\textwidth]{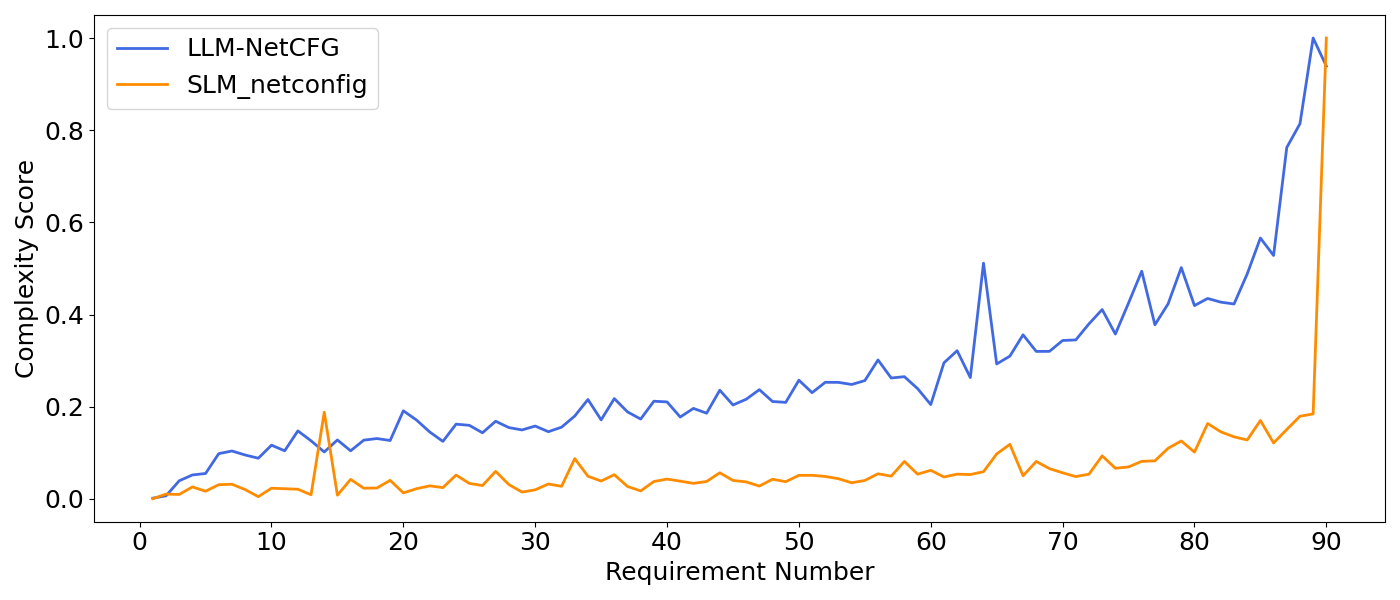}
\label{fig_Complexity}
\end{figure}

Figure~\ref{fig_Complexity} presents the complexity levels of the configurations produced by LLM-NetCFG and SLM\_netconfig, operating with the fine-tuned small question-configuration model, on the 99-sample dataset. LLM-NetCFG’s outputs exhibited broad complexity, often falling in the medium-to-high range (0.3–0.8 and above), indicating that it frequently generates unnecessarily intricate configuration command sets due to limited contextual information. In contrast, SLM\_netconfig consistently generated more concise configurations, with most results concentrated in the 0.1–0.2 range and only one outlier at 1.0. The outlier in SLM netconfig, with a complexity of 1.0, occurs because that specific sample required an unusually detailed set of commands to meet its configuration needs. Unlike most cases where the model produces concise outputs, this instance reflects either the inherent complexity of the scenario or a rare deviation from the patterns learned during fine-tuning, showing that even optimized models can generate more extensive configurations when necessary. These results highlight the effectiveness of fine-tuning carried out in SLM\_netconfig for producing optimized configurations that maintain functionality while minimizing unnecessary command redundancy.

\begin{figure}[h!]
\centering
\caption{Translation Time per Requirement - LLM-NetCFG vs SLM\_netconfig.}
\includegraphics[width=0.49\textwidth]{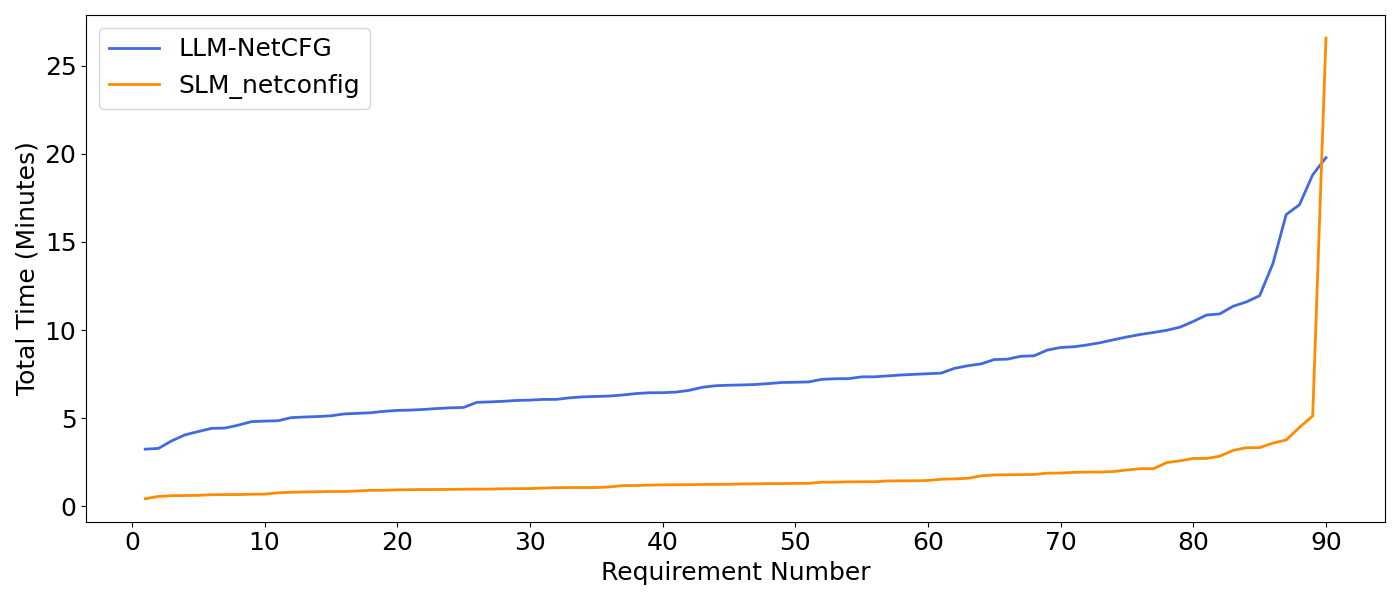}
\label{fig_Totaltime}
\end{figure}

Figure~\ref{fig_Totaltime} illustrates the translation time obtained for LLM-NetCFG and SLM\_netconfig. LLM-NetCFG required 5 to 7.5 minutes for configurations with complexities between 0.15 and 0.25, from 7.5 to 12 minutes for complexities between 0.3 and 0.5, and from 17 to 20 minutes for highly complex configurations (when complexity approached 1.0). This upward trend indicates that our previous solution becomes less efficient as configuration complexity increases. Conversely, SLM\_netconfig obtained significantly superior performance, translating most requirements in under 5 minutes, with the majority completed in just 1–3 minutes. Only one high-complexity case required 26 minutes; despite this outlier, the overall trend indicates that SLM\_netconfig is faster than LLM-NetCFG, reflecting the advantages of its optimized architecture and domain-specific fine-tuning.

\subsection{Highlights}
\label{highlights}

The observed success of SLM\_netconfig across the evaluations can be directly linked to the fine-tuning process and the specific design of the question-configuration dataset. Fine-tuning allowed the model to internalize domain-specific syntax, hierarchical command structures, and semantic dependencies unique to Cisco IOS configurations, capabilities that general-purpose pretrained models typically lack. This specialization enabled the model to reduce redundancy, generate syntactically coherent commands, and ensure that each configuration step aligned precisely with the intended operational goal. In contrast, LLM-NetCFG retained broad linguistic competence but lacked the specialized reasoning required for precise network-configuration synthesis.

In terms of syntax, SLM\_netconfig benefited from exposure to structured examples where each command corresponded to a specific functional requirement, reinforcing disciplined keyword usage, parameter selection, and command ordering. Without such domain adaptation, LLM-NetCFG frequently diverged from strict syntax, even when producing grammatically coherent text. Thus, fine-tuning was essential for transitioning from general linguistic capability to technical correctness.

Regarding configuration goal accuracy, SLM\_netconfig achieved superior results due to the question-driven dataset design, which aligned each input requirement with an explicit target configuration. This mapping encouraged the model to interpret requirements through a goal-oriented reasoning process rather than shallow text-to-command translation, yielding a deeper understanding of operational intent and reducing errors from ambiguous phrasing. LLM-NetCFG lacked such structured alignment, leading to frequent omissions of supporting commands or incomplete implementation of configuration objectives. Thus, the fine-tuned model's success in achieving goals underscores the effectiveness of training data that mirrors the real-world reasoning chains used by network engineers.

The differences between LLM-NetCFG and SLM\_netconfig also become evident in complexity and translation time results. SLM\_netconfig fine-tuned internal representations allow it to identify the minimal command set required to satisfy each goal, avoiding redundant or obsolete configurations. This compression effect not only simplifies the output but also accelerates generation, as the model no longer expends effort on irrelevant or contradictory instructions. In contrast, the non-fine-tuned LLM-NetCFG frequently generates verbose or overlapping command sequences, increasing computational cost and response latency. Fine-tuning thus enables more efficient inference by streamlining decision pathways and reducing token-level uncertainty.

\section{Conclusion and Future Work}
This work presented SLM\_netconfig, a fine-tuned small language model designed for autonomous network configuration generation. Across all evaluation metrics, the proposed model consistently outperformed the baseline LLM-NetCFG. SLM\_netconfig produced more precise and goal-aligned configurations, exhibited lower structural complexity, and achieved substantially faster translation times. Collectively, these results confirm the effectiveness of domain-specific fine-tuning and show that small models, when trained with curated configuration datasets, can rival or surpass larger LLMs while offering advantages in efficiency, privacy, and deployability on local infrastructure. Despite these improvements, the partially correct and incorrect cases allowed identifying potential avenues for further optimization. Enhancing training with stricter structural constraints, integrating self-verification or consistency-checking mechanisms, and adopting RAG could help mitigate hallucinations and strengthen command validity. Additionally, more comprehensive training data—such as multi-device configuration workflows and expanded support for multiple network vendors—would improve model adaptability in heterogeneous operational environments.

Future work will investigate deployment in larger-scale operational settings featuring diverse network topologies and will extend model capabilities beyond Cisco IOS to multi-vendor ecosystems. Further research will also explore automated repair mechanisms for partially correct outputs and closed-loop feedback systems that integrate device-level validation into the inference process. These directions aim to advance towards a fully autonomous, robust, verifiable, and scalable SLM-based framework for network self-configuration.

\printbibliography

\vfill\eject
\vspace{-1cm}
\begin{IEEEbiographynophoto}
{Oscar G. Lira} (GS'17) is pursuing his Ph.D degree in Computer Science at the State University of Campinas, Campinas, Brazil, and M.Sc in Computer networks and computer security and his degree in Telecommunications engineering in University Galileo of Guatemala, Guatemala. His research interests include network and service management, network virtualization, machine learning, and large language models.
\end{IEEEbiographynophoto}
\vspace{-1.3cm}
\begin{IEEEbiographynophoto}
{Oscar M. Caicedo}(GS'11--M'15--SM'20) is a full professor at the Universidad del Cauca, Colombia, where he is a member of the Telematics Engineering Group. He received his Ph.D. degree in computer science (2015) from the Federal University of Rio Grande do Sul, Brazil, and his M.Sc. in Telematics Engineering (2006) and his degree in Electronics and Telecommunications Engineering (2001) from the Universidad del Cauca. His recent research interests include network and service management, network functions virtualization, software-defined networking, machine learning and large language models for networking, and network softwarization. He serves as Series Editor for the IEEE Communications Magazine Series on Network Softwarization and Management and Associate Editor of the IEEE Networking Letters. He served and keeps serving as Technical Program Co-Chair for IEEE-sponsored international workshops and conferences. 
\end{IEEEbiographynophoto}
\vspace{-1.3cm}
\begin{IEEEbiographynophoto}
{Nelson L. S. da Fonseca} (M'88--SM'01) received the Ph.D. degree in computer engineering from the University of Southern California, Los Angeles, CA, USA, in 1994. He is currently a Full Professor with the Institute of Computing, State University of Campinas, Campinas, Brazil. He has authored or co-authored more than 450 papers and supervised over 80 graduate students. Dr. da Fonseca is currently ComSoc President-Elect. He served as VP Conferences, VP of Technical and Educational Activities VP of Publications, VP Member Relations, the Director of Conference Development, Latin America Region, and On-Line Services. He is the former Editor-in-Chief of the IEEE COMMUNICATIONS SURVEYS AND TUTORIALS. He was a recipient of the 2023 ComSoc Education Award, 2020 IEEE ComSoc Harold Sobol Award for Exemplary Service to Meetings \& Conferences, 2012 IEEE ComSoc Joseph LoCicero Award for Exemplary Service to Publications, the Medal of the Chancellor of the University of Pisa in 2007, and the Elsevier Computer Networks Journal Editor of Year 2001 Award.
\end{IEEEbiographynophoto}
\end{document}